\documentclass[12pt,twoside]{article}
\newcommand{\dessin}[4]{%
\vspace*{20pt}%
{\hfill\strut \mbox{\includegraphics[scale=#1]{#2}}\strut\hfill}%
\vspace*{5pt}\\ %
{\hfill\strut{\hbox{\vbox{{\bf Fig. #3}\ \sl #4}}\strut}\hfill\strut}%
\strut\vspace{10pt}\\ }%
\usepackage{graphicx}
\begin{document}
{\hfill UMH-MG 01-01}\vspace{4cm}
\begin{center}

{\Large\bf Scalar perturbations in a\\[5pt] primordial inflationary
scenario\\[5pt]}
\medskip

J\'ulio C. Fabris\footnote{e-mail: fabris@cce.ufes.br}\medskip

Departamento de F\'{\i}sica, Universidade Federal do Esp\'{\i}rito Santo,
29060-900, Vit\'oria, Esp\'{\i}rito Santo, Brazil

and\\
Philippe Spindel\footnote{spindel@umh.ac.be}\medskip

M\'ecanique et Gravitation, Universit\'e de Mons-Hainaut, 7000 Mons,
Belgium\medskip

\end{center}

\begin{abstract}
We compute the spectral index for scalar perturbations
generated in a primordial inflationary model.
In this model, the transition
of the inflationary phase to the radiative era is achieved through the
decay of the cosmological term leading a
second order phase transition and
the characteristics of the model allow to implement a set
of initial conditions
where the perturbations
display a thermal spectrum when they emerge from the horizon.
The obtained value for the spectral index is equal to 2,
a result that depends very weakly on the various parameters of the model
and on the initial conditions used.
\vspace{0.7cm}

PACS number(s): 04.20.Cv., 04.20.Me
\end{abstract}

\section{Introduction}
The big-bang standard model, based on the
Friedmann-Lema\^{\i}tre-Robertson-Walker
  (FLRW) solutions of Einstein's
equations \cite{lfrw}, suffers from some
drawbacks which leaded cosmologists to propose the so-called
inflationary models \cite{GB}. The common element to all
these models is an exponential expansion phase of the size
of the Universe prior to the beginning of the standard scenario that describes
the adiabatic era extending from temperatures
of order of $10^{-4}$ at Grand Unification era (GU) until $10^{-32}$,
today's temperature, in Planck's units (which
are the units  that will be used throughout all this work).
\par
 From a
kinetic
point of view, the exponential expansion phase corresponds to a de Sitter
geometry \cite{HE} which later on will transform itself into  a FLRW geometry.
 From a physical point of view, this scenario raises (at least) two
questions. What is the nature of the effective cosmological
constant during the de Sitter phase of the history of the Universe? What
is
the mechanism which leads to the end of this phase? Among the different
inflationary models, the first one to be proposed \cite{BEG}
looks specially interesting. It consists of a primordial inflationary
model describing the birth and grow of the Universe as the result of a
cooperative
mechanism of zero total energy, generated by a quantum fluctuation occurring
in a flat background. The idea that
a quantum fluctuation may be the initial seed for the birth of the
Universe
has first been proposed by Tryon in 1973 \cite{TR}. It is because
the total hamiltonian
of gravity coupled to matter is zero (an important consequence of the
invariance of general relativity with respect to time reparametrisation)
and the energy contribution of the gravitational conformal modes is negative
  that it is possible
to obtain
non-trivial semi-classical solutions of Einstein equations where 
gravity and matter emerges
from an empty flat space in a cooperative process.
\par
In the model proposed by Brout et al. \cite{BEG}, matter is described
phenomenologically
through the action of a scalar field non-minimally coupled to gravity.
A consistent asymptotic solution of the field equations \cite{BES,
BEFGNTS} describes a
de Sitter geometry where the role of the cosmological constant is
played by the (constant) density of the created matter which has,
due to the energy conservation law,
the characteristic equation of state of a cosmological constant
$\rho_\Lambda + p_\Lambda = 0$. On the other hand,
the consistency of the model implies that the (effective) cosmological
constant $\Lambda \equiv 8\pi\rho_\Lambda$ is a free parameter of the obtained
solutions.
It must not necessarily be of order one, what justify a posteriori the
semi-classical treatment employed.
However, the mass of the scalar quanta fields is very large \cite{BES}:
$M \simeq 6\sqrt{\pi}$, this value being almost insensible to the value
of the cosmological constant \cite{Sp} even when the vacuum
polarization effects
are taken into account \cite{SV}; these effects
play no significative role in the domain of validity
of the model ($\Lambda << 1$). On the other hand, the high value
of the  mass of the quanta is the signal that gravitational strong coupling
must play an important role in the physics of this problem\cite{En}. The scalar
field used in the model must be interpreted as a
phenomenological description of complex objects, like black holes
\cite{CE}.
This interpretation, by the way, is confirmed by the analysis of the
renormalization of the gravitational constant \cite{En}. Indeed,
the renormalized gravitational constant becomes singular \cite{DC}
exactly for
the critical value of the mass field $M = 6\sqrt{\pi}$, what is the origin
of the analytical
solution \cite{Sp2} of the semi-classical equations which describes, for this
special value, the transition from a flat space to a
singularity. Since the cosmological ``constant'' in this model is
described  a gas  of black holes, their subsequent evaporation through the
Hawking radiation mechanism  may imply a time-decreasing of this effective
cosmological "constant", generating hot radiation in the Universe.
\par
For the moment, there is no consensus about how the inflationary
phase ends. However, it is possible to distinguish, from the
phenomenological
point of view, essentially two types of transition to an adiabatic era:
a first order phase transition, employed in the so-called old and
new inflationary models, which deals with an effective cosmological
constant given by
the energy of a scalar field;
a second order phase transition, as it be discussed here,
based on a variable cosmological "constant".
\par
The evaporation of the black holes or the infra-red fluctuations of the
gravitational conformal modes \cite{AM} being responsible for the
instability of
the de Sitter geometry, it seems important to extract observational
consequences of the model described above through the analysis of
a simple scenario. In \cite{BS} analytical solutions
of the FLRW type for the phenomenological model described above
were obtained for any value of the spatial curvature
($k = 0, \pm 1$). They depend on four physical parameters:
the initial value of the cosmological constant, $\Lambda_0$, its
residual value $\Lambda_\infty$, the initial density of radiation
in the Universe $\sigma_0$ and the characteristic time of transmutation
of the cosmological constant into radiation energy $\tau$.
The choice of a decaying of the cosmological constant into pure
radiation is motivated in two ways.
 From the physical point of view, the temperatures encountered in this model
permit us to ignore
the masses of the usual degrees of freedom. From the mathematical
point of view, the background model is exactly integrable and their
perturbations easily computable.
\par
Our purpose in the present work is to study the evolution of scalar
perturbations in the inflationary model
described before. We will restrict ourselves to the case
$\Lambda_\infty = \sigma_0 = 0 $ and $k = 0$.
These hypothesis lead
to the advantage that approximate solutions to the perturbed equations
can be easily obtained without to oversimplify the physics.
Our model differs essentially in two ways from the usual inflationary
models. First (after having fixed the residual gauge freedom allowed by
the coordinate transformations), the perturbation equations in the 
framework of this two fluid model
(the effective cosmological constant and radiation) will result in a
third order perturbed equation (analytically solvable in the radiation regime
and that otherwise can very well be approximated
by a $2 + 1$ system of  decoupled equations, also  analytically solvable).
The second characteristic
of the model we discuss here concerns the nature of the initial spectrum
of the fluctuations. As it will be verified later, the fluctuations
evolve freely, i.e., they are determined by the Einstein's equations,
only
from the moment where the temperature of the created radiation is
significative. Consequently, we will adopt as initial fluctuation
spectrum
a thermal spectrum, instead of a (hypothetical) spectrum
inspired by quantum fluctuations of a vacuum state (a choice
imposed more by metaphysical\footnote{During the
fifties, Bertrand Russell observed: ``The accusation of metaphysics has
become in philosophy something like being a security risk in the
public service. \ldots The only definition I have found that fits
all cases is: `a philosophical opinion not held by the present
author'." \cite{Ru}.} than by physical considerations; see
however the discussions in \cite{MRS, MBNP}, for instance).
\par
This paper is organized as follows. In section 2, the field equations
and the solution for the background are reminded. In section 3
the perturbed quantities are settled out, and approximate solutions
presented. In section 4, the spectrum
of perturbations is discussed. Section 5 is devoted to the conclusions.

\section{The field equations}

In order to be complete, let us review briefly the solutions that will
be used in our perturbative study. The geometry of the Universe is
supposed
to be of the type flat FLRW:
\begin{equation}
ds^2 = - dt^2 + a^2(t)(dx^2 + dy^2 + dz^2) \quad .
\end{equation}
This metric is a reasonable simplification
for the primordial stages of
the Universe. Moreover, in our case, due to the existence of an
inflationary
phase, the curvature can also be neglected in later stages of the
evolution
of the Universe.
The unperturbed material source of the
Einstein's equations is the sum of two comoving fluids:
\begin{equation}
T^\mu_\nu = (\stackrel{0}{\rho} + \stackrel{0}{p})\stackrel{0}{{u}^\mu}
{\stackrel{0}{u}}_\nu + \stackrel{0}{p}\delta^\mu_\nu\equiv T^\mu_{(rad.)\nu}+
T^\mu_{(cos.)\nu}
\end{equation}
with $\stackrel{0}{\rho} = \rho_{(rad.)} + \rho_\Lambda$, 
$\stackrel{0}{p} = p_{(rad.)}
+ p_\Lambda$.
The terms representing the radiation perfect fluid ($\rho_{(rad.)}$, 
$p_{(rad.)}$) are
supposed to represent the high energy matter
degrees of freedom
  under the form of radiation:
\begin{equation}
\rho_{(rad.)} = \nu\frac{\pi^2}{30}T^4
\end{equation}
where $T$ is the temperature and $\nu$ is the effective number of
degrees of freedom that we will take as being those of the standard model
($\nu = 106.75$), this value of $\nu$ constituting, of course, a lower bound
at the energy considered.
The term $\rho_\Lambda = {\Lambda(t)}/{8\pi}= - p_\Lambda$ is purely
phenomenological. It represents the
density of black holes created during the cosmogenesis phase
and that evaporate later.We will assume it to be of the form~:
\begin{equation}
\Lambda(t) = \Lambda_0e^{-t/\tau} \qquad ,
\end{equation}
where the parameter
$\tau$ is the characteristic time
of the black hole evaporation which is of the order of the
lifetime of the created black holes~:
\begin{equation}
\tau \sim\frac{2560\pi}{\nu}M^3_{bh} \approx 10^5 \quad.
\end{equation}
The equation of state of radiation, ${\rho_{(rad.)}}={3}\/ p_{(rad.)}$, permits
to obtain a linear second order differential equation
for the
variable $Z = a^2$ (a dot representing a derivative with respect to the
cosmological time $t$):
\begin{equation}
\ddot Z - \frac{4}{3}\Lambda_0e^{-t/\tau}Z = 0 \quad .
\end{equation}
The solution of this equation is given by a combination of modified
Bessel functions:
\begin{equation}
\label{bs}
Z = \alpha K_0(\varpi e^{-\xi}) + \beta I_0(\varpi e^{-\xi}) \quad ,
\end{equation}
where $\xi = \frac{t}{2\tau}$ and $\varpi = \sqrt{\frac{16}{3}\tau^2
\Lambda_0}$. The coefficients $\alpha$ and $\beta$ are given by
\begin{equation}
\label{ic}
\alpha = {\varpi}(I'_0(\varpi) + I_0(\varpi)) \quad ,
\quad \beta = - {\varpi}(K'_0(\varpi) + K_0(\varpi)) \quad
,\end{equation}
and are fixed by the conditions $Z(0) = 1$ and $\dot Z(0) =
+\sqrt{4\Lambda_0/3}$.
\par
Notice that there was an error of sign in the
ref. \cite{SV}. The scale factor written there corresponds
to $\dot Z < 0$, and the plateau appearing in the graph of the
entropy as function of time is due to the crossing of the singularity
$Z = 0$. The graphs for the entropy and temperature as functions
of time, for initial
conditions (\ref{ic}) corresponding to an expanding Universe, are
shown on figures 1 and 2.\\
  \begin{minipage}{7cm}
\dessin{1}{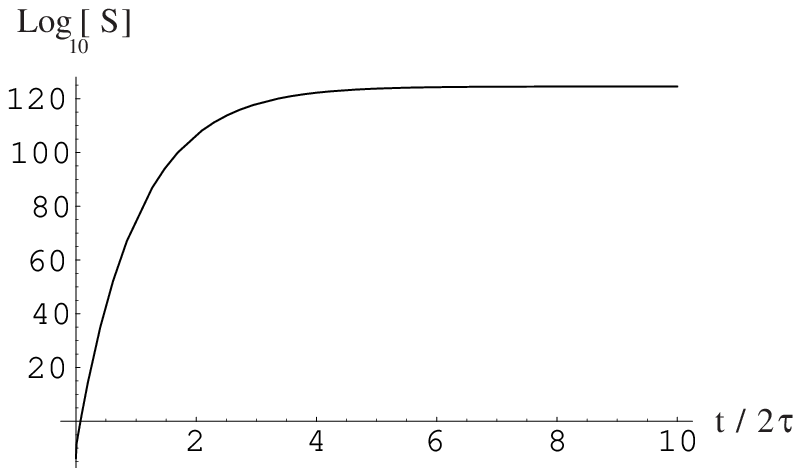}{1}{ Logarithmic plot of the radiation entropy as
function of time for $\Lambda_0=5.5 10^{-7}$ i.e. $N_e=80$.}
\end{minipage} \hfill
\begin{minipage}{7cm}
\dessin{1}{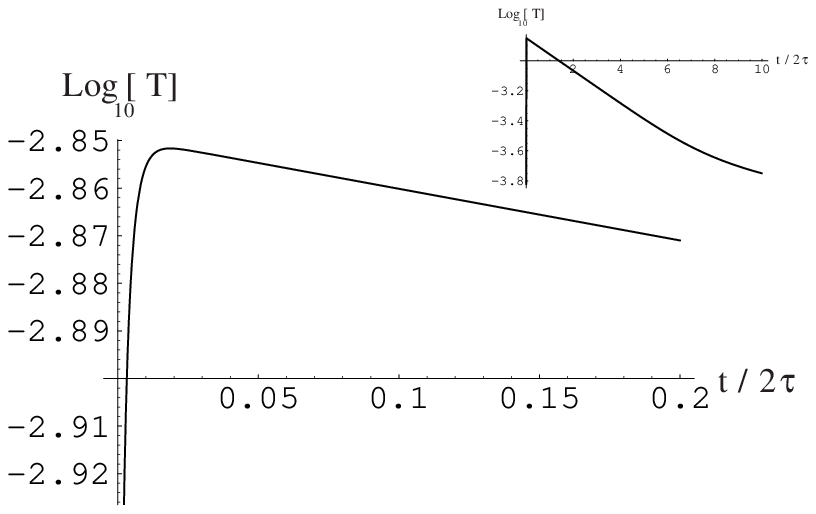}{2}{ Logarithmic plots of the radiation temperature as
function of time for $\Lambda_0=5.5 10^{-7}$ i.e. $N_e=80$.}
\end{minipage}
\section{The perturbed equations}

The perturbed equations for the scalar modes around the background
solutions exhibited above are obtained through the standard methods.
We suppose the metric perturbations
\begin{equation}g_{\mu\nu}(t, \vec x) = \stackrel{0}{g}_{\mu\nu}(t) +
\stackrel{1}{g}_{\mu\nu}(t, \vec x)
\end{equation}
to be synchronous, i.e., such that $\stackrel{1}{g}_{\mu0} = 0$. The
perturbations
in density and velocity read
\begin{equation}
\rho(t,\vec x) = \stackrel{0}{\rho}(t) + \stackrel{1}{\rho}(t,\vec x)
\quad ,
\quad u^\mu(t,\vec x) = \stackrel{0}{{u}\strut^\mu} +
{\stackrel{1}{u}}\strut^\mu(t,\vec x) \quad ,\end{equation}
with ${\stackrel{0}{u}}\strut^\mu = (1,0,0,0)$.
All perturbed quantities can be expressed through a
Fourier decomposition
\begin{eqnarray}
\stackrel{1}{g}_{ij}(t,\vec x) &=& \frac{1}{\pi^{3/2}}\int h_{ij}(t,\vec
n)
e^{-i\vec n.\vec x}d^3\vec n \quad ,\\
\stackrel{1}{\rho}(t,\vec x) &=& \frac{1}{\pi^{3/2}}\int \delta \rho(t,\vec n)
e^{-i\vec n.\vec x}d^3\vec n \quad \quad ,\\
{\stackrel{1}{u}}_\mu(t,\vec x) &=&
\frac{1}{\pi^{3/3}}\int{\stackrel{1}{u}}_\mu(t,\vec n)e^{-i\vec n.\vec
x}d^3\vec n \quad ,
\end{eqnarray}
with $\vec n.\vec x = \sum_i n^ix^i$, $\vec n$ being the
comoving wave vector
of the perturbations.
\par
Introducing those perturbed quantities
into Einstein's equations, we obtain three coupled perturbed equations:
\begin{eqnarray}
\label{ep1}
({a^2\dot h})^. = a^2\delta\rho \quad ,\\
\label{ep2}
(a^4\delta\rho)^. + (\rho + p)a^4\theta = 8\pi(\rho + p)a^4h \quad ,\\
\label{ep3}
((\rho + p)a^5\theta)^. = \frac{n^2}{3}a^3\delta\rho \quad ,
\end{eqnarray}
where $\theta = \vec n.\vec u$ and $h = - 16\pi a^{-2}\sum_i h_{ii}$.
The system of equations (\ref{ep1},\ref{ep2},\ref{ep3}) is
underdetermined since $\rho = \rho_{(rad.)} + \rho_\Lambda$ and
consequently $\delta\rho = \delta\rho_{(rad.)} + \delta\rho_\Lambda$.
We need a specific equation for the effective cosmological constant.
Hence, we  assume
\begin{equation}
\biggr(\nabla_\mu T^{\mu\nu}_{(cos.)}\biggl)u_\nu \equiv -
\nabla_\nu(\rho_\Lambda
u^\nu)
- p_\Lambda\nabla_\nu u^\nu = \tau^{-1}\rho_\Lambda
\end{equation}
which reduces to $u^\nu\partial_\nu\rho_\Lambda = \tau^{-1}\rho_\Lambda$, since
$\rho_\Lambda = - p_\Lambda = \Lambda(t)/8\pi$. For adiabatic
perturbations ($\delta\rho_\Lambda + \delta p_\Lambda = 0$), it results
($\tau$ being supposed to be a constant given by the microphysics)
\begin{equation}
{\stackrel{1}{u}}\strut^\mu\partial_\mu\stackrel{0}{\rho}_\Lambda +
\partial_t\stackrel{1}{\rho}_\Lambda = \tau^{-1}\stackrel{1}{\rho}_\Lambda
\quad .
\end{equation}
Since ${\stackrel{1}{u}}\strut^0 = 0$,  we obtain $\stackrel{1}{\rho}_\Lambda =
L(\vec x)e^{-t/\tau}$.
But this solution is a coordinate artifact.
Indeed, the coordinate conditions $g_{\mu0} = 0$ yet remain satisfied
when the
residual
infinitesimal transformations generated by the vector fields
\begin{equation}
\epsilon_0 = \phi(\vec x) \quad , \quad \epsilon_i = \psi_i(\vec x)a^2 -
\partial_i\phi(\vec x)a^2\int \frac{dt}{a^2}
\end{equation}
are performed. The choice $\phi(\vec x) = \frac{\tau L(\vec 
x)}{\Lambda_0}$ allows to
make $\delta\Lambda(t) = 0$. Notice that, in this
way, the equations (\ref{ep1},\ref{ep2},\ref{ep3}) are now gauge
invariants
with respect to the remaining coordinate
transformations.
Indeed, residual transformations $\epsilon_0 = 0$, $\epsilon_i =
\psi_ia^2$
only modify $h_{ij}(t,\vec x)$ up to a term whose time dependence is
given by $a^2$; hence, $h$ is defined up to a constant, leaving
$\dot h$ invariant.
In other words, in the synchronous gauge $g_{0\mu} = 0$, $\delta\rho_{\Lambda}$ may
be fixed to zero and
all solutions of
the perturbed equations (\ref{ep1},\ref{ep2},\ref{ep3}) now acquire
a physical meaning, in contrast to what happens in the framework of
  a one-fluid model.
In this later case, the third order
perturbed equation may be reduced, using the residual coordinate
freedom, to a second order equation. Here, we face a third order equation
whose three integration constants have physical meaning.
\par
 From now on, we introduce the new set of variables
\begin{equation}
{\cal T} = (\rho_{(rad.)} + p_{(rad.)})a^5\theta \quad , \quad
{\cal R} = a^4\delta\rho_{(rad.)} \quad , \quad
{\cal H} = a^2\dot h \quad ,
\end{equation}
which permit us to write the system of equations
(\ref{ep1},\ref{ep2},\ref{ep3}) as
\begin{eqnarray}
\label{ep1'}
\dot{\cal H} &=& \frac{{\cal R}}{a^2} \quad ,\\
\label{ep2'}
\dot{\cal R} &=& - \frac{{\cal T}}{a} + 8\pi(\rho_{(rad.)} +
p_{(rad.)})a^2{\cal H} \quad , \\
\label{ep3'}
\dot{\cal T} &=& \frac{n^2}{3}\frac{{\cal R}}{a} \quad .
\end{eqnarray}
The solutions of this system and their cosmological implication will be the
subject of the next section.
\section{Evolution of perturbations}

We are not interested here in the general solution of the  system of equation
(\ref{ep1'}-\ref{ep3'}) governing the evolution of the perturbations,
for arbitrary values of their parameters. We shall only focus on the
values of the
parameters that are relevant for the cosmological model we consider here.
In order that an inflationary model is acceptable, the scale factor
must be amplified at least by $70$ e-folds, before the Universe
reaches the GU temperature $T_{GU} = 10^{-4}$ where
radiative era takes place. At this moment,
the expansion of the Universe is described by the asymptotic
expression
\begin{equation}
Z_{as} = \alpha\frac{t - t_s}{2\tau} \equiv \alpha(\xi - \xi_s)
\quad , \quad \xi_s \equiv \frac{t_s}{2\tau} \simeq \ln\frac{\varpi}{2} +
\gamma \quad ,
\end{equation}
$\gamma \approx 0.577$ being the Euler-Mascheroni constant.
Using the Einstein's equation $G_{00} = 8\pi T_{00}$, we obtain
\begin{equation}
\xi_{GU} - \xi_s = \biggr(\frac{128\pi^3\tau^2}{90}\nu
T^4_{GU}\biggl)^{-1/2}
\end{equation}
Hence, $\alpha \sim \biggr(\frac{128\pi^3\tau^2}{90}\nu
T^4_{GU}\biggl)^{1/2}
e^{2N_e} \approx 10^{62}$ and $\beta \approx 10^{-65}$ for a e-fold number
$N_e = 75$. Accordingly, unless for $\xi$ very close to zero, we may
neglect the second term in (\ref{bs}).
 From the asymptotical expression of the $K_0$ Bessel function
appearing in eq.(\ref{ic}) we may deduce the value of $\varpi$
(i.e. $\Lambda_0$) insuring a given number $N_e$ of e-fold~:
\begin{equation}
\varpi = 2N_e + \frac{1}{2}\ln\frac{2^{12}\pi^2\nu T^4_{GU}}{3}
+ \frac{1}{2}\ln\biggr[2N_e + \frac{1}{2}\ln\frac{2^{12}\pi^2 \nu
T^4_{GU}}{3}\biggl] \quad .\label{varpi}
\end{equation}
This implies $\Lambda_0 << 1$ for a large range of reasonable values of
$N_e$, in particular $\Lambda_0 \approx 1.4\times10^{-7}$ for $N_e = 75$,
what is compatible with the domain of validity of the model, while
for $\Lambda_0 = 1/(192\pi^2)$, which corresponds to the assumption of
thermal equilibrium between de Sitter temperature and
Hawking temperature, we obtain $N_e \simeq 2400$.
\par
According to the range of the values settled by the initial
conditions, we may build approximate solutions of the perturbed equations.
For instance, let us suppose that ${\cal R}$ scales with $\alpha$ as
$\alpha^K$. From the perturbed equations, using the conservation of the
entropy of the radiation as the cosmological constant goes to zero,
we obtain ${\cal H} \sim \alpha^{K-1}$ and
${\cal T} \sim \alpha^{K - 1/2}$.
From the perturbed equations, it comes out
that in this case $\dot{\cal R} \sim \frac{{\cal T}}{a} \sim
\alpha^{K -1}$, which is negligible compared to
$\frac{32\pi}{3}\rho_{(rad.)}a^2{\cal H}
\sim \alpha^K$. Hence, the perturbed equations can be approximated by~:
\begin{eqnarray}
\label{as1}
\dot{\cal H} &=& \frac{{\cal R}}{a^2} \quad ,\\
\label{as2}
\dot{\cal R} &=& \frac{32\pi}{3}\rho_{(rad.)}a^2{\cal H} \quad , \\
\label{as3}
\dot{\cal T} &=& \frac{n^2}{3}\frac{{\cal R}}{a} \quad ,
\end{eqnarray}
which can be integrated by quadratures. The general solution
is~:
\begin{eqnarray}
\label{solap1H}
{\cal H} &=& {\cal H}_i\frac{Z(t_i)}{Z(t)} +
\frac{\mu}{Z(t)}\int_{t_i}^tZ(t')dt'
\quad , \\
\label{solap1R}
{\cal R}(t) &=&  Z(t)\dot{\cal H}(t) =
\mu Z(t) - \dot Z(t){\cal H}(t) \quad ,\\
{\cal T} &=& {\cal T}_i + \frac{n^2}3\int_{t_i}^t\frac{{\cal
R}(t')}{a(t')}dt'
\end{eqnarray}
with
\begin{equation}
\mu = \frac{1}{Z(t_i)}\biggr({\cal R}_i + {\dot Z}(t_i){\cal H}_i
\biggl)\label{mu}
\end{equation}
and ${\cal R}_i$, ${\cal H}_i$, ${\cal T}_i$ being the initial values of
${\cal R}$,
${\cal H}$ and ${\cal T}$.\\
On the other hand, if ${\rho a^3{\cal H}}$ can be neglected in front of
${\cal T}$, the system reduces
to
\begin{equation}
\frac{d{\cal R}}{d\zeta} = - {\cal T} \quad ,
\quad \frac{d{\cal T}}{d\zeta}
= \frac{n^2}{3}{\cal R} \quad , \quad \dot{\cal H} = \frac{{\cal R}}{a^2}
\end{equation}
where $\zeta = \int_{t_i}^t \frac{dt'}{a(t')}$. Its general solution
is then
\begin{eqnarray}
{\cal R} &=& {\cal R}_i\cos\biggr(\frac{n}{\sqrt{3}}\zeta\biggl)
- \frac{\sqrt{3}}{n}{\cal T}_i \sin\biggr(\frac{n}{\sqrt{3}}\zeta\biggl)
\quad ,\label{solap2R}\\
{\cal T} &=& {\cal T}_i\cos\biggr(\frac{n}{\sqrt{3}}\zeta\biggl) +
\frac{n}{\sqrt{3}}{\cal R}_i\sin\biggr(\frac{n}{\sqrt{3}}\zeta\biggl)
\quad ,\label{solap2T}\\
{\cal H} &=& {\cal H}_i + \int_{t_i}^t\frac{{\cal R}(t')}{\tau Z(t')}dt'
\quad \label{solap2H}.
\end{eqnarray}
\par
The initial time $t_i$, where the initial values of ${\cal R}$,
${\cal T}$ and
${\cal H}$ have to be considered, depend on the size of the perturbations.
Today the temperature is of the order of $T_0 \sim 2\times10^{-32}$, the
cosmological observational data which are somehow free of
astrophysical noises mainly concerns sizes $\lambda_0$ from
$100 Mpc$ up to $3.000 Mpc$, i.e., $2\times10^{62}l_{Pl}$ up to
$6\times10^{63}l_{Pl}$. At these scales the Universe can
be considered as homogeneous. At the grand unification epoch, the size
of these
fluctuations are shortened by a factor
${T_0}/{T_{GU}} = 2\times10^{-28}$.
They must still be reduced by a factor
${a(t_i)}/{a(t_{GU})}$ in order to obtain their values at
the initial time. Moreover, the evolution of the fluctuations
is given by equations (\ref{ep1},\ref{ep2},\ref{ep3}) only
once they grow faster than the microscopic interactions can propagate. It is
only
then that their behavior is no more dictated by the microphysics but
only by the overall cosmological expansion.
This implies that the evolution of
perturbations with comoving wave numbers $n$ and present day size $\lambda_0$
will only be determined by the perturbation equations from an initial
time $t_i(n)$ given by $\dot a(t_i(n)) = n$, i.e.,
\begin{equation}
\label{it}
\dot a(t_i(n)) = \frac{2\pi}{\lambda_0}\frac{T_{GU}}{T_0}e^{N_e}\quad .
\end{equation}
\par
This time $t_i(n)$ may be easily estimated by noting
that $\varpi e^{-\xi_{GU}}$, and thus $\varpi e^{-\xi_{n}}$
(where $\xi_i(n) = t_i(n)/2\tau$),
depends very slowly on the number of e-folds. Moreover, near $t_i(n) = 0$
equation (\ref{it}) easily may be  linearized. Thus for each value of
$\lambda_{\star}$, it is  possible to determine a minimal number
$N^{min}(\lambda_{\star})$ of e-folds such that fluctuations of size
as $\lambda_{\star}$ today emerges from the horizon at $t = 0$:
\begin{equation}
N^{min}(\lambda_{\star}) =
\ln\biggr[\frac{\lambda_{\star}}{4\pi\tau}\frac{T_0}{T_{GU}}
\ln\frac{\lambda_{\star} T_{0}}{4\pi^2\tau T_{GU}\sqrt{\xi_{GU}}}\biggl] + 1.93
\quad ,
\end{equation}
and giving this number, we obtain from eq. (\ref{varpi}), for a given number
  $N_e$ of e-folds,
\begin{equation}
t_i(\lambda_{\star}) \approx
2\tau\ln\biggr[\frac{\varpi(N_e)}{\varpi(N^{min})}\biggl]
\end{equation}
as the time where a fluctuation of size $\lambda_{\star}$ has
emerged from the horizon.
We restrict ourselves to positive values of $t_i$(we do not know
the physics before $t < 0$). This implies that the minimum number
of e-fold
between $t = 0$ and the beginning of the grand unification era we will
consider is at least
$75$.
\par
The decay of the cosmological constant increases very
rapidly the temperature (see fig. 1),
but nevertheless we shall suppose that the density fluctuations are given
by the laws of the equilibrium statistical mechanics \cite{LL5}:
\begin{equation}
\delta\rho(t_i,n) =
\nu_{eff}^{1/2}T^{5/2}\biggr(\frac{a(t_i(n))}{n}\biggl)^{3/2}
\frac{1}{a^3(t_i(n))}\qquad .
\end{equation}
The extra factor $a^{-3}(t_i(n))$ is the jacobian that we have to introduce
when we consider the Fourier transforms with respect to
the comoving variables $n$ which are related to the physical momentum
$n_{phys}$ at time $t_i$ by
\begin{equation}
n_{phys} = \frac{n}{a(t_i)} = \frac{2\pi}{\lambda_{\star}}\frac{T_{GU}}{T_0}
\frac{e^{N_e}}{a(t_i)} \quad .
\end{equation}
For consistency, we may check that the ratio of the proper energy
density of such fluctuation to the temperature $\epsilon(n)/T_i(n)$ is
approximatively equals to $1.8$ when the fluctuation emerges from the
horizon, the temperature $T_i(n)$ at this moment being of the order
or $10^{-3}$. We may also check that these values are very few sensitive
to the values of $\tau$ and $N_e$. Moreover, we also assume the initial
spectrum of $\theta(t_i(n))$ and $\dot h(t_i(n))$ being given by classical
physics, i.e., they are not affected by the overall expansion of the
Universe. Using standard arguments \cite{LL5,LL6}, we obtain
\begin{eqnarray}
\theta(t_i,n) &\sim& \nu_{eff}^{1/2}T_i^{-3/2}(n)n^{-1/2}a(t_i(n))^{-5/2}
\quad ,\\
\dot h(t_i,n) &\sim& \nu_{eff}^{1/2}T_i^{5/2}(n)n^{-5/2}a(t_i(n))^{-1/2}
\quad .
\end{eqnarray}
Hence, ${\cal R}(t_i,n) = Z^2(t_i(n))\delta\rho(t_i,n)$, ${\cal T}(t_i,n) \sim
n{\cal R}(t_i,n)$ and
${\cal H}(t_i,n) \sim {{\cal R}(t_i,n)}/{na(t_i,n)}$
The second relation means that the longitudinal mode $h(t_i,p)$
is determined by the gravitational fluctuations,
with a characteristic time
of the order of the size of the fluctuations considered (of course,
one also could ask about the metaphysical content of these considerations,
that we adopt in the absence of a better scenario).
\par
With these assumptions, it is possible to verify that at initial
times ${\cal T}/a$ is about $2-5\times10^3$ times greater than
$\frac{32\pi}{3}\rho_{(rad.)}a^2{\cal H}$~:
\begin{equation}
\frac{32\pi}{3}\rho_ia_i^3\frac{{\cal H}_i}{{\cal T}_i} \approx
8\pi\frac{a^2(t_i)\rho_{(rad.)}(t_i)}{n^2}
\approx \frac{3}{16\tau^2n^2}\varpi e^{-\xi_i/2}e^{\varpi(1 - e^{-\xi_i})}
<< 1 \quad ,
\end{equation}
a result again almost insensitive to the values of $N_e$ and $\tau$.
As a consequence, during  a first short period of time we may
approximate the
system of the perturbation equations by the eqs
(\ref{as1},\ref{as2},\ref{as3}). Hence, one  verify that the
functions ${\cal H}$, ${\cal R}$ and ${\cal T}$ given by eqs
(\ref{solap2R},\ref{solap2T},\ref{solap2H}) change little in the
beginning of the evolution of perturbations. But
the ratio between the two terms in the right hand
side of (\ref{ep2}) behaves as~:
\begin{equation}
\frac{32\pi}{3}\rho_{(rad.)}a^3\frac{{\cal H}}{{\cal T}} \sim
8\pi \frac{\rho_{(rad.)}(t)a^3(t)}{n^2a(t_i)}
\end{equation}
and increases very quickly. Hence, for values of $\xi$ very near
$\xi_i$, the system will be in the domain of validity of the
first approximate solution (\ref{solap1H},\ref{solap1R}).
Taking into account the uncertainties on the the parameters
of the problem (initial spectrum, precise duration of validity
of the perturbed equations, etc.), we can adopt them as  solutions of
the perturbed equations.
 From these considerations, it comes out that at the grand unification
temperature, the density contrast $\delta_{GU} = |\delta\rho(t_{GU})/\rho_{GU}|$ is
given by
\begin{equation}
\delta_{GU} =  \frac{{\cal
R}(t_{GU})}{\rho_{GU}Z^2_{GU}} \approx \frac{32\pi}{3}\frac{\mu
e^{2N_e}}{\alpha^2} \quad ,
\end{equation}
and all dependence on the wave number $n$ is contained in the
value of the integration constant $\mu$. This constant  is
given by the sum of two terms (see eq. (\ref{mu}), but the first one
$Z(t_i)\delta\rho_i$ is significantly more
important than the second one.
Hence, the spectrum index depends only on the initial spectrum of
density perturbations and on the dates where, for a given size 
$\lambda_0$, the fluctuations emerge from
the horizon. As the temperature $T(t_i,n)$ is almost constant during 
the interval of time where the relevant perturbations 
($100Mpc<\lambda_0<3000Mpc$) cross the horizon while
$a(t_i(n)) \div n$  we immediately obtain the spectral index
\begin{equation}
n_s = 4 + \frac{d}{d\ln n}\ln[(\delta\rho(n)/\rho)^2] \approx 2
\quad .
\end{equation}

\section{Conclusions}

In this work, we have studied an inflationary model that joins
the radiative regime by a second order phase transition.
This model is inspired in an inflationary primordial scenario where
the transition from a de Sitter phase to a radiative phase occurs
due to the evaporation of primordial black holes.
It must be considered as a source of a possible alternative to the usual
inflationary
models, and not as a definitive model.
Indeed, the underlined physics leading to the background geometry is
phenomenological and asks for more deep fundamental investigations.
It offers, however, the conceptual advantage of making no appeal to
a self-interacting scalar field.
The kinetics of the phase transition asks also for a more rigorous
argumentation.
However, we believe our results not to be dependent on the details
of the of the model as the spectral index obtained is very little
sensible to the values of $N_e$ and $\tau$.
\par
Moreover, we have introduced hypothesis on the nature of the initial
spectrum of the fluctuations.
In particular, we employed a thermal spectrum for the fluctuations
as they emerge from horizon. Notice that, from the conceptual point of
view, the spectrum more delicate is
that of the $h$ but this quantity
does not play a significative role in the computation of the spectral
index,
since the gravitational fluctuations is dominated by those, much more
important, fluctuations of matter, which is governed by ${\cal R}$.
The nature of the spectrum of $\delta\rho$ affects the spectral index
essentially by the function $a(t_i) \sim n$ which appears in the
jacobian of density fluctuation defining $\delta\rho_i$ and on the
behavior of $Z(t_i,n)$, as appears in the expression for the
initial conditions. Of course the  obtained value of the spectral index
is not in agreement with observational data. However here we have only
discussed  scalar fluctuations; it remains to see if tensorial 
fluctuations cannot reconcile the model with the expected 
Harrison-Zeldovitch spectra.
But such an analysis reopen the question of the choice of the initial
spectrum of the gravitational perturbations, a question whose answer, 
in our opinion is beyond our present knowledge of physics.

\section*{ACKNOWLEDGMENTS}
The authors thank J\'er\^ome Martin for fruitful discussions
and CAPES (Brazil) for
partial financial support. Ph. S. thanks Robert Brout,
Fran\c{c}ois Englert and Marianne Rooman for numerous enlightening discussions about many
aspects of the study performed here, Jean Bricmont for noticing him 
ref. \cite{Ru} and CAPES (Brazil) for partial
financial support. He also thanks the Departamento de F\'{\i}sica
of UFES for its warm hospitality.

\end{document}